\documentclass[prd,onecolumn,preprintnumbers,showpacs,nofootinbib,notitlepage,floatfix]{revtex4-1}

\usepackage{graphicx}
\usepackage{amsmath}
\usepackage{amssymb}
\usepackage{slashed}
\usepackage{amsfonts}
\usepackage{xcolor}
\usepackage{multirow}
\usepackage{array}
\usepackage{mathrsfs}
\usepackage{CJK}
\usepackage{hyperref}

\newcolumntype{C}{c|}

\allowdisplaybreaks

\begin{document}

\title{The Electromagnetic Form Factors of Pseudoscalar Mesons within the Light-Front Quark Model}

\author{Shuai Xu$^{a}$}
\email{xushuai@zknu.edu.cn}
\author{Xiao-Nan Li$^{b}$}
\email{lixn@tlu.edu.cn}
\author{Xing-Gang Wu$^{c}$}
\email{wuxg@cqu.edu.cn}

\affiliation{
$^a$School of Physics and Telecommunications Engineering, Zhoukou Normal University, Henan 466001, P.R. China\\
$^b$School of Electrical Engineering, Tongling University, Anhui 244000, P.R. China\\
$^c$Department of Physics, Chongqing Key Laboratory for Strongly Coupled Physics, Chongqing University, Chongqing 401331, P.R. China}

\date{\today}

\begin{abstract}
	
In this paper, we investigate the electromagnetic form factors (EMFFs) and charge radii of pseudoscalar mesons within the light-front quark model (LFQM). Using parameters derived from the confinement of mesonic decay constants, we obtain numerical results, which indicate the following: (i) The EMFFs of charged and neutral mesons exhibit significant differences in their endpoint behaviors but show similar asymptotic behavior in the high momentum transfer ($Q^2$) regions. For the EMFFs of light mesons such as $\pi$ and $K^+$, our results are in excellent agreement with experimental data in the low $Q^2$ regions. For the charge radii of mesons, our results also show rough consistency with predictions from other approaches. (ii) For charged mesons, the peak values of $Q^2 F_P(Q^2)$ are approximately proportional to the mass difference $\Delta m$ between their constituent quarks. Moreover, the mean square radii $\left\langle r^2 \right\rangle_P$ of charged mesons decrease with increasing meson mass and decreasing $\Delta m$. For neutral mesons, their charge radii are primarily determined by the electric charge of the heavy quark. These results indicate that quark mass asymmetry significantly influences the behavior of the EMFFs and charge radii of mesons. Experimental data to test these predictions would thus be of great interest.

\end{abstract}

\maketitle

\section{Introduction}

A fundamental challenge in hadron physics is elucidating the internal structure of hadrons in terms of their constituent quarks and gluons. The electromagnetic form factors (EMFFs) of mesons serve as essential quantities in this endeavor, describing the spatial distribution of charge and current within the hadron. These form factors provide crucial information about hadronic structure and the dynamics governing constituent interactions. Pseudoscalar mesons represent particularly accessible systems for such studies. For a spinless pseudo-scalar ($P$) particle, the electromagnetic structure is parameterized by a single form factor, $F_{P}(Q^2)$, offering a straightforward yet powerful probe for achieving this goal.

Experimentally, the EMFFs of the $\pi$ and $K^+$ mesons at low momentum transfers $Q^2<0.4~\text{GeV}^2$ have been precisely measured at Fermilab~\cite{Dally:1981ur, Dally:1982zk, Dally:1980dj} and CERN~\cite{Amendolia:1984nz, NA7:1986vav, Amendolia:1986ui}. Measurements at moderate momentum transfers $Q^2\sim[2,6]~\text{GeV}^2$ are being pursued using the upgraded JLab accelerator~\cite{Arrington:2021alx, Dudek:2012vr}. While the EMFF of the pion can be directly measured from elastic pion-electron scattering in the low $ Q^2 $ regions, extracting it at intermediate and high $ Q^2 $ becomes challenging due to the pion's short lifetime, which limits its use as a stable target in such experiments. The challenging regime of intermediate momentum transfers, extending up to $Q^2 \sim 10~\text{GeV}^2$, has been explored at Cornell~\cite{Bebek:1974ww, Bebek:1976qm, Bebek:1977pe}, DESY~\cite{Kunszt:1979ur, Ackermann:1977rp}, and JLab~\cite{JeffersonLabFpi:2000nlc, JeffersonLabFpi-2:2006ysh, JeffersonLabFpi:2007vir, JeffersonLab:2008gyl, JeffersonLab:2008jve, Carmignotto:2018uqj}. Precise data at higher momentum transfers $10~\text{GeV}^2 < Q^2 < 40~\text{GeV}^2$ are anticipated from the forthcoming EIC experiment at BNL~\cite{AbdulKhalek:2021gbh}. However, despite their potential to reveal more information about internal structure and bound-state dynamics due to higher flavor asymmetry, the EMFFs of heavy mesons remain largely unexplored experimentally.

Theoretically, numerous approaches have been applied to compute these quantities, including Lattice QCD (LQCD)~\cite{Li:2017eic, Ding:2024lfj}, QCD Sum Rules (QCDSR)~\cite{Nesterenko:1982gc}, the Bethe-Salpeter Equations~\cite{Xu:2019ilh, Roberts:1994dr, Maris:1999nt, Qin:2011dd, Xu:2020loz, Maris:1997tm, Qin:2019oar, Xu:2024fun}, Constituent Quark Models (CQMs)~\cite{Moita:2021xcd}, the Extended Nambu-Jona-Lasinio Model (ENJL)~\cite{Luan:2015goa}, the Contact Interaction Model (CI)~\cite{Hernandez-Pinto:2023yin}, the Algebraic Model (AM)~\cite{Almeida-Zamora:2023bqb}, and others~\cite{Huang:2004fn, Huang:2004su, Wu:2008yr, Wang:2019mhm, Das:2016rio, Aliev:2019lsd, Simonis:2016pnh, Bose:1980vy, Lahde:2002wj}. Leveraging the advantages of manifest Lorentz invariance and conceptual simplicity, the relativistic light-front quark model (LFQM) has been quite successful in analyzing various hadronic form factors, decay constants, distribution amplitudes, and so on~\cite{Brodsky:1992px, Jaus:1999zv, Grach:1983hd, Chung:1988my, Hwang:2001th, Chang:2019obq, Chang:2018zjq, Chang:2019mmh, Jaus:1989au, Jaus:1996np, Cheng:2003sm, Choi:2013mda, S:2024adt, Choi:2017uos,Xu:2025aow}. In this work, we utilize the LFQM framework to study the EMFFs of pseudoscalar mesons.

The paper is organized as follows. Section~\ref{sec:2} provides a brief review of the theoretical formalism for the EMFF and the LFQM. Numerical results and discussion are presented in Section~\ref{sec:3}. Finally, conclusions are given in Section~\ref{sec:4}.

\section{Theoretical Formulation} \label{sec:2}

The definition of the pseudoscalar mesonic EMFF is
\begin{eqnarray}
\langle M(p'')|j^{\mu}_{\text{em}}|M(p')\rangle = P^\mu F_{P}(Q^2),
\label{eq:1}
\end{eqnarray}
where $p'{(p'')}$ is the momentum of the initial (final) bound state, $P^{\mu}=p'^{\mu}+p''^{\mu}$ and $q^{\mu}=p'^{\mu}-p''^{\mu}$. In this paper the Drell-Yan-West frame $q^{+}=0$ is chosen, where the momentum transfer $Q^2=-q^2=\mathbf{q}^2_{\perp}$ lies in the space-like region. The charge current is approximated as a one-point function
\begin{eqnarray}
j^{\mu}_{\text{em}}(0) = \sum_{q,\bar{q}} e_q \bar{q}(0)\gamma^{\mu}q(0),
\label{eq:2}
\end{eqnarray}
where $e_u=2/3$ and $e_d=-1/3$ are the electric charges of the up and down quarks, respectively.

It is straightly to extract $F_{P}(Q^2)$ from Eq.\eqref{eq:1} within the LFQM, more details are given in Refs.~\cite{Choi:1997iq, Choi:1999nu}. And we obtain
\begin{equation}
F_{P}(Q^2)=e_{q}I(m_q,m_{\bar{q}},Q^2)+e_{\bar{q}}I(m_{\bar{q}},m_q,Q^2),
\label{eq:3}
\end{equation}
where
\begin{equation}
I(m_q,m_{\bar{q}},Q^2)=\int_{0}^{1} dx \int d^2\mathbf{k}_{\perp}
\sqrt{\frac{\partial k_z}{\partial x}} \phi_{R}(x, \mathbf{k}_{\perp})
\sqrt{\frac{\partial k'_z}{\partial x}}\phi_{R}(x, \mathbf{k'}_{\perp})\frac{[xm_q+\bar{x}m_{\bar{q}}]^2+ \mathbf{k}_{\perp} \cdot \mathbf{k'}_{\perp}}
{\sqrt{[xm_q+\bar{x}m_{\bar{q}}]^2 + \mathbf{k}^2_\perp} + \sqrt{[xm_q+\bar{x}m_{\bar{q}}]^2 + \mathbf{k'}^2_\perp}},
\end{equation}
where ${\bf{k'}}_{\perp}={\bf{k}}_{\perp}+\bar{x}{\bf{q}}_{\perp}$, $\bar{x}=1-x$ and the longitudinal momentum $k_z=(x-\frac{1}{2})M_0+\frac{m_q^2-m_{\bar{q}}^2}{2M_0}$ in the Jacobian $\frac{\partial k_z}{\partial x}$ of the variable transformation $\{x,{\bf{k}}_{\perp}\}\rightarrow \{{\bf{k}}_{\perp},k_z\}$. The invariant mass of bound state $M_0$ is defined as
\begin{eqnarray}
M_0^2= \frac{m_{q}^2+{\bf{k}}_{\perp}^2}{\bar{x}}+\frac{m_{\bar{q}}^2+{\bf{k}}_{\perp}^2}{x},
\label{eq:4}
\end{eqnarray}
which plays a significant role in analyses of covariance and self-consistency of the standard and covariant approaches of LFQM~\cite{Jaus:1999zv, Cheng:2003sm, Choi:2013mda,Xu:2025aow}. In this work, a Gaussian-type radial wave function $\phi_R$ is taken, e.g.
\begin{eqnarray}
\phi_R(x,{\bf{k}}_{\perp})= \frac{4\pi^{3/4}}{\beta^{3/2}} \mathrm{exp}(-\frac{{\bf{k}}_{\perp}^2+k_z^2}{2\beta^2}), \label{eq:5}
\end{eqnarray}
where the parameter $\beta$ is related to the size of bound state and will be discussed in the next section.

\section{Numerical Results and Discussion} \label{sec:3}

The model parameters for constituent quark masses and Gaussian parameter $\beta$ are listed in Tables~\ref{quarkmass} and ~\ref{beita}, which are mainly obtained by fitting the world averages of experimental data on mesonic decay constants from the Particle Data Group (PDG)~\cite{ParticleDataGroup:2022pth}. For comparison, we also present the parameters in the previous works~\cite{Hwang:2001th, Chang:2019obq, Choi:2009ai, Hwang:2010hw, Choi:1997iq, Cheng:2003sm, Verma:2011yw, Choi:2015ywa}.

\begin{table*}[htb]
\begin{center}
\begin{tabular}{ccccc}
\hline\hline
&$m_q$~&~~$m_s$~&~~$m_c$ ~&~~$m_b$\\
\hline
Ref.\cite{Chang:2019obq, Choi:2009ai, Hwang:2010hw, Cheng:2003sm, Verma:2011yw, Choi:2015ywa}& [0.21,0.26] & [0.37,0.48]  & [1.38,1.80] & [4.64,5.20]  \\
This Work & $0.25(\pm0.01)$ & $0.50(\pm0.02)$ & $1.80(\pm0.09)$ & $5.10(\pm0.25)$ \\
\hline \hline
\end{tabular}
\caption{The values of quark masses (in units of GeV) suggested in the previous works and fitted in this work, where $q = u,d$.}
\label{quarkmass}
\end{center}
\end{table*}


\begin{table*}[htb]
\begin{center}
\begin{tabular}{cccccccc}
\hline\hline
&$\beta_{q\bar{q}}$&$\beta_{q\bar{s}}$&$\beta_{q\bar{c}}$&$\beta_{s\bar{c}}$&$\beta_{q\bar{b}}$&$\beta_{s\bar{b}}$&$\beta_{c\bar{b}}$\\
\hline
Ref.\cite{Hwang:2001th, Chang:2019obq, Choi:1997iq} & [0.306,0.366] & [0.303,0.389] & [0.416,0.468] & [0.450,0.549] &[0.477,0.531] & [0.485,0.578] & [0.813,0.896] \\
This Work & $0.321(\pm0.016)$ & $0.352(\pm0.017)$ & $0.465(\pm0.023)$ & $0.522(\pm0.026)$ & $0.535(\pm0.026)$ & $0.594(\pm0.03)$ & $0.883(\pm0.044)$\\
\hline\hline
\end{tabular}
\caption{The values of Gaussian parameter $\beta$ (in units of GeV) suggested in the previous works and fitted in this work, where $q = u,d$.}
\label{beita}
\end{center}
\end{table*}

\subsection{The Pions and Kaons}\label{sec:pika}

\begin{figure*}[htb]
\centering
\includegraphics[width=8.1cm,height=7cm]{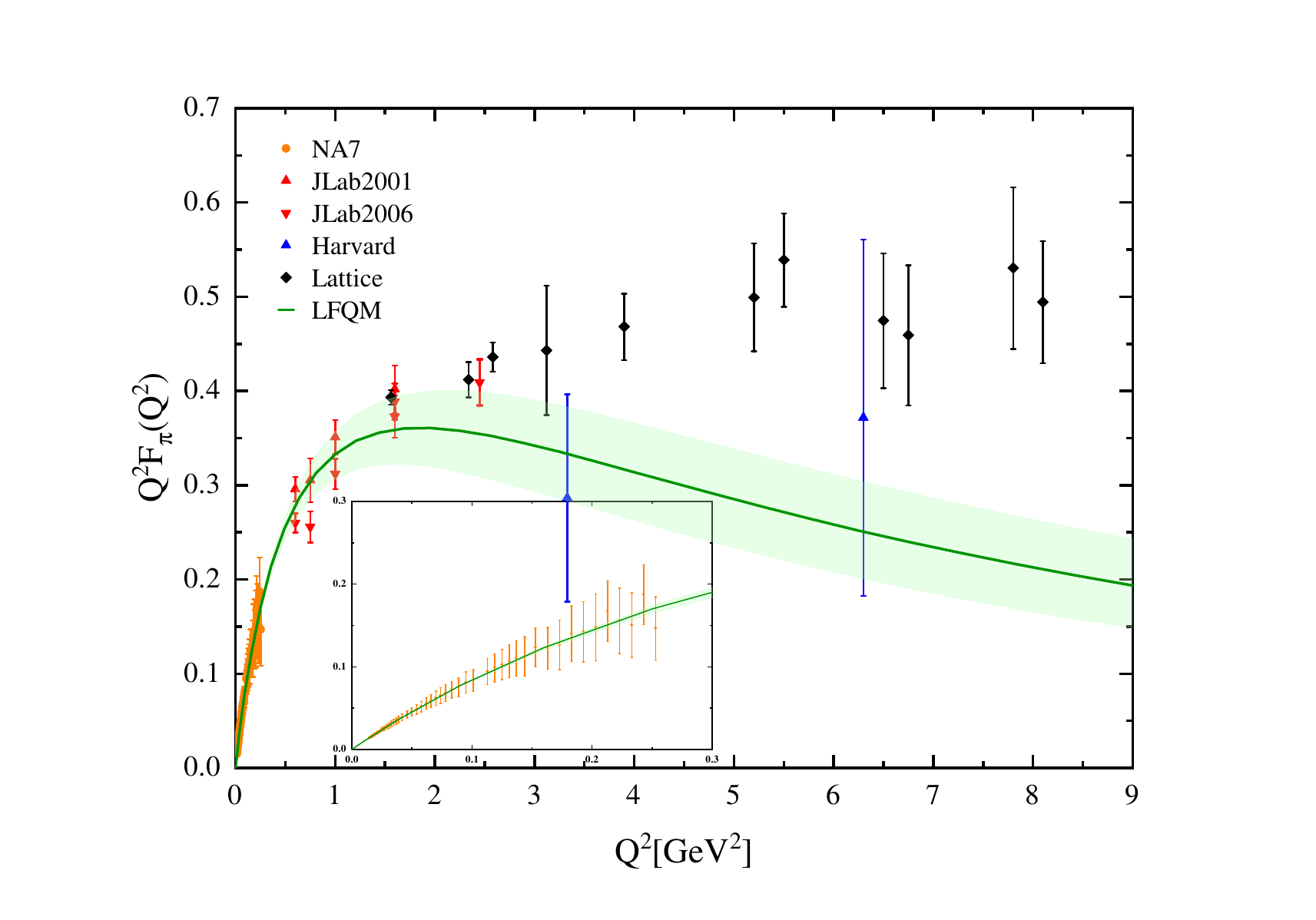}
\includegraphics[width=8.1cm,height=7cm]{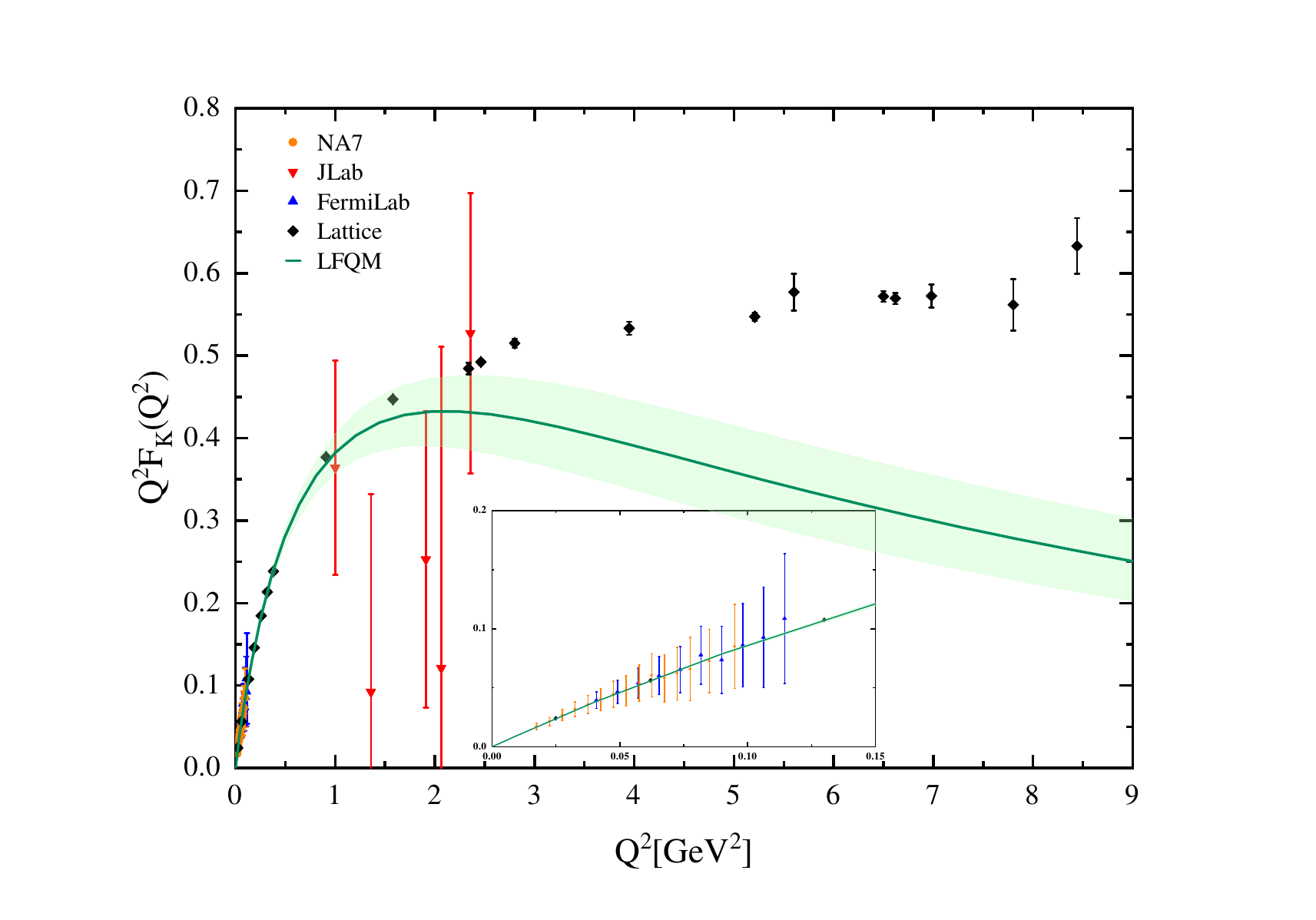}
\caption{Left: The EMFF of the charged pion obtained from the LFQM, compared with experimental data from Refs.~\cite{NA7:1986vav, JeffersonLabFpi:2000nlc, JeffersonLabFpi-2:2006ysh, JeffersonLabFpi:2007vir, Bebek:1977pe} and lattice data from Ref.~\cite{Ding:2024lfj}. Right: Same as the left panel, but for the charged kaon, with experimental data from Refs.~\cite{Amendolia:1986ui, Carmignotto:2018uqj, Dally:1980dj} and lattice data from Ref.~\cite{Ding:2024lfj}.}
\label{PI}
\end{figure*}

We compare our results for the EMFFs $Q^2F_{\pi^+}(Q^2)$ and $Q^2F_{K^+}(Q^2)$ of the charged pion and kaon with the measured data from Refs.~\cite{NA7:1986vav, JeffersonLabFpi:2000nlc, JeffersonLabFpi-2:2006ysh, JeffersonLabFpi:2007vir, Bebek:1977pe, Amendolia:1986ui, Carmignotto:2018uqj, Dally:1980dj}, respectively. Some comments are provided as follows:
\begin{itemize}
\item In the low momentum transfer regions $Q^2<0.3$ GeV$^2$, our LFQM results for $Q^2F_{\pi^+}(Q^2)$ and $Q^2F_{K^+}(Q^2)$ are in good agreement with the data from NA7~\cite{NA7:1986vav}, even through both theoretical uncertainty and experimental error are very small, as shown in Fig.~\ref{PI}. This indicate that the endpoint behavior of EMFFs for light mesons, which encode the information of internal charge distribution, are well described and grasped. The experimental data in regions $Q^2\sim[0.3,0.5]$ GeV$^2$ is lack and expected in the future experiments.

\item In the intermediate momentum transfer regions $Q^2\sim[0.5,2]$ GeV$^2$, for charged pion, our results are agreement with the experimental data from NA7~\cite{JeffersonLabFpi:2000nlc, JeffersonLabFpi-2:2006ysh, JeffersonLabFpi:2007vir}, Harvard~\cite{Bebek:1977pe} and the recent outcomes of LQCD~\cite{Ding:2024lfj}. Similarly, our theoretical results for the charged kaon are also in good agreement with the experimental data from Refs.~\cite{Amendolia:1986ui, Carmignotto:2018uqj, Dally:1980dj} and LQCD results from Ref.~\cite{Ding:2024lfj}.

\item In the high momentum transfer regions $Q^2>2$ GeV$^2$, our theoretical predictions show obvious asymptotic behavior earlier than LQCD~\cite{Ding:2024lfj}~\footnote{Such a discrepancy may be explained by considering the higher-twist structures~\cite{Huang:2004su}; this analysis is in preparation.}. Based on the analysis of perturbative QCD~\cite{Lepage:1979zb}, the asymptotic behavior of $F_{P}(Q^2)$ should be $1/Q^2$ and $Q^2F_{\pi}(Q^2)=8\pi \alpha_s(Q^2)f_{\pi}^2$ for $Q^2\to \infty$ is checked. The forthcoming experiments at JLab~\cite{JeffersonLab:2008jve} may bring a criterion.
\end{itemize}

\begin{figure*}[htb]
\centering
\includegraphics[width=8.1cm,height=7cm]{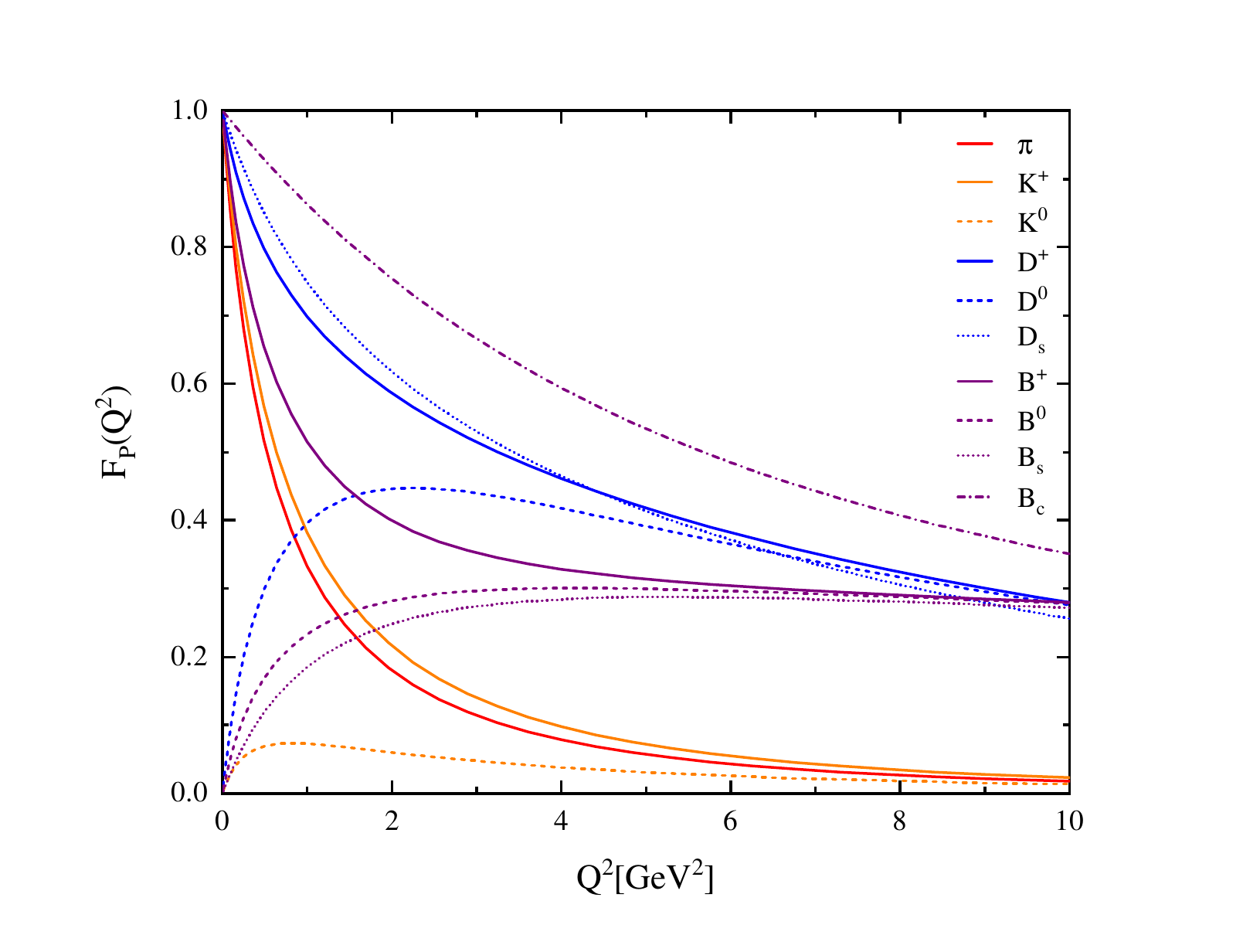}
\includegraphics[width=8.1cm,height=7cm]{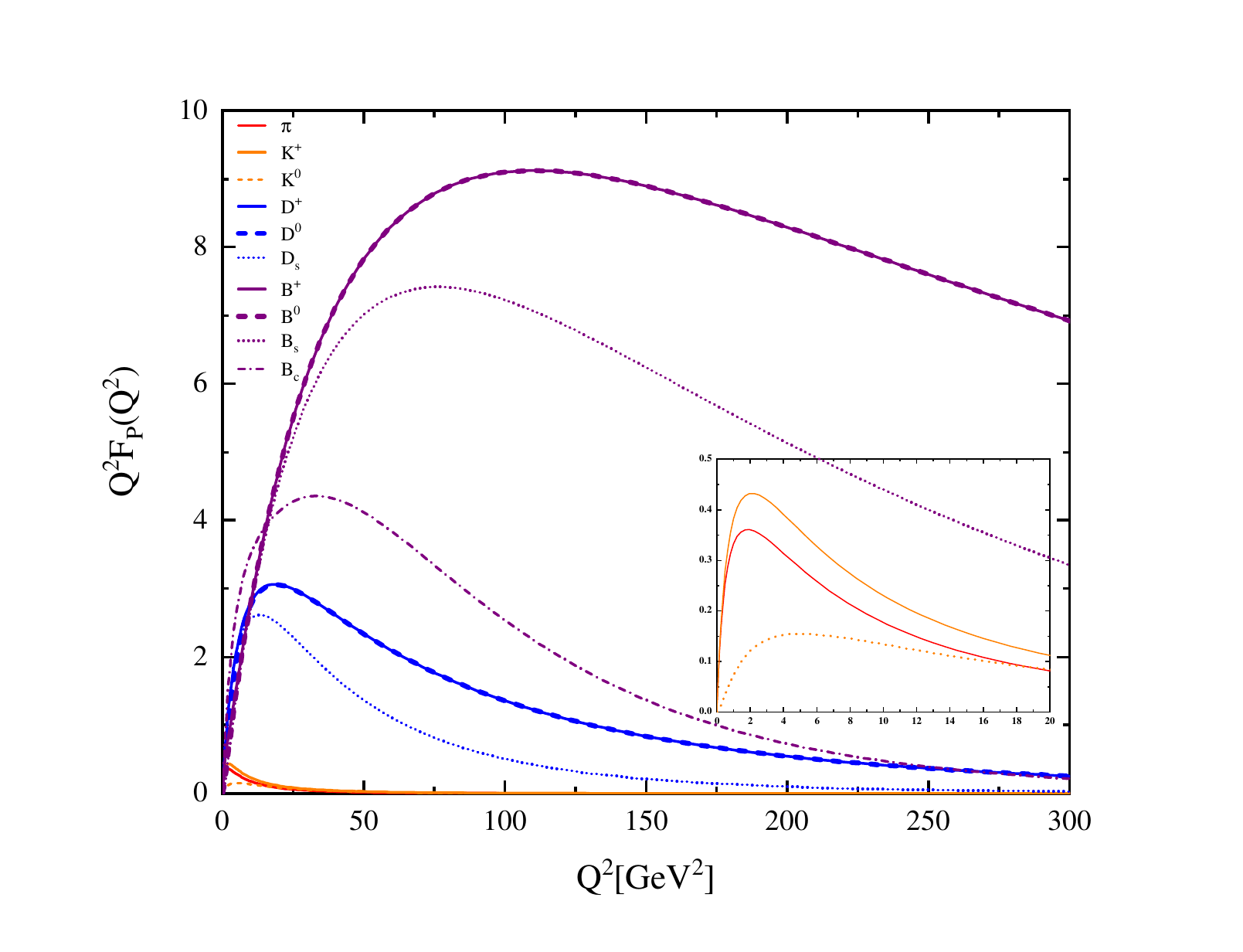}
\centering
\caption{$Q^2$-dependence of the EMFFs $F_{P}(Q^2)$ and $Q^2F_{P}(Q^2)$ for pseudoscalar mesons.}
\label{ff10}
\end{figure*}

Principally, the EMFFs are the same for the charged and neutral pion under isospin symmetry, which treats the $u$ and $d$ quarks are identical~\footnote{Since the charged pion ($\pi^{\pm}$) is a $u\bar{d}$ or $\bar{u}d$ bound state and the neutral pion ($\pi^0$) is a mixture of $u\bar{u}$ and $d\bar{d}$, their EMFFs are not exactly identical. However, isospin symmetry implies they should be very similar, and differences are small due to mild isospin breaking.}. However, due to SU(3)-flavor breaking effects, the EMFFs are differ in the kaonic case. We present the EMFFs $F_{K^0}(Q^2)$ and $Q^2 F_{K^0}(Q^2)$ of the neutral kaon in Fig.~\ref{ff10}. The static property $F_P(0)=e_P$, i.e., $F_{K^+}(0)=1$ and $F_{K^0}(0)=0$ is exhibited. This relationship also holds for the heavy mesons discussed subsequently.

\begin{table*}[h]
\begin{center}
\begin{tabular}{ccCcccccccc}
\hline\hline
Meson&$\Delta m$(GeV) & $Q^2F_{P}(Q^2)$&$\sqrt{\langle r^2\rangle_P}$(fm) & BSEs~\cite{Xu:2024fun} & CQM~\cite{Moita:2021xcd} & ENJL~\cite{Luan:2015goa} & CI~\cite{Hernandez-Pinto:2023yin} & AM~\cite{Almeida-Zamora:2023bqb} & LQCD~\cite{Li:2017eic} & PDG~\cite{ParticleDataGroup:2022pth}\\
\hline
$\pi^{+}(u\bar{d})$ & 0 &0.361 & $0.668^{+0.035}_{-0.031}$ & 0.646 &0.665&0.57&0.45&-& 0.648(15) & 0.659(4) \\
$K^{+}(u\bar{s})$ & 0.25 & 0.432 & $0.610^{+0.020}_{-0.018}$ & 0.608 &0.551&0.54&0.42&-& - & 0.560(31) \\
$K^{0}(d\bar{s})$ & 0.25 &0.154 &${0.302 i}^{+0.022 i}_{-0.021 i}$ & 0.253$i$ &-&-&-&-& - & 0.277(18)$i$ \\
$D^{+}(c\bar{d})$ & 1.55 &3.064 & $0.411^{+0.015}_{-0.015}$ & 0.435 &0.505&0.46&-&0.680& 0.450(24) & - \\
$D^{0}(c\bar{u})$ & 1.55 &3.050 & ${0.534 i}^{+0.026 i}_{-0.024 i}$ & 0.556$i$ &-&-&0.36$i$&-& - & - \\
$D_{s}(c\bar{s})$ & 1.30 &2.615 & $0.301^{+0.011}_{-0.010}$ & 0.352 &0.377&0.39&0.26&0.372& 0.465(57) & - \\
$B^{+}(u\bar{b})$ & 4.85 &9.122 & $0.564^{+0.022}_{-0.022}$ & 0.619 &-&0.74&0.34&0.926& - & - \\
$B^{0}(d\bar{b})$ & 4.85 & 9.122 &${0.396 i}^{+0.016 i}_{-0.016 i}$ & 0.435$i$&-&-&-& - &- \\
$B_{s}(s\bar{b})$ & 4.60 & 7.418 & ${0.281 i}^{+0.013 i}_{-0.012 i}$ & 0.337$i$ &-&-&0.24$i$&0.345$i$& - & - \\
$B_{c}(c\bar{b})$ & 3.30 & 4.355& $0.189^{+0.010}_{-0.009}$ & 0.219 &-&-&0.17&0.217& - & - \\
\hline \hline
\end{tabular}
\caption{Numerical results for the charge radii of various pseudoscalar mesons. The second column lists the constituent quark mass differences, the third column shows the peak values of $Q^2F_{P}(Q^2)$, and the fourth column presents the charge radii of the mesons calculated in this work. Subsequent columns provide charge radii from other theoretical approaches and the PDG world average, respectively.}
\label{aaa}
\end{center}
\end{table*}

The mean square charge radius of meson is determined by the slope of $F_{P}(Q^2)$ at $Q^2=0$
\begin{eqnarray}
\left\langle r^2 \right\rangle_P=-\left. 6\frac{d F_P\left(Q^2\right)}{d Q^2}\right|_{Q^2=0}.
\label{eq:6}
\end{eqnarray}
Numerical results for the charge radii of various pseudoscalar mesons are presented in Table~\ref{aaa}. As shown in Table~\ref{aaa}, for charged mesons, our calculated charge radii are consistently below 1 fm, a characteristic scale of nucleons. The charge radius of the pion, $\sqrt{\left\langle r^2 \right\rangle_{\pi}}=0.668(31)$ fm, is in excellent agreement with the PDG average of $0.659(4)$ fm~\cite{ParticleDataGroup:2022pth}. Meanwhile, the kaon's charge radius, $\sqrt{\left\langle r^2 \right\rangle_{K^+}}=0.610(18)$ fm, is slightly larger than the PDG average of $0.560(31)$ fm~\cite{ParticleDataGroup:2022pth}. It is noticed our result of $K^+$ is consistent with the measurement from BLATNIK data $0.620\pm0.037 $ fm \cite{ParticleDataGroup:2022pth}.

Fig.~\ref{ff10} shows in the low $Q^2$ region (near $Q^2=0$), the slopes of the EMFFs $F_P(Q^2)$ for all neutral pseudoscalar mesons are positive, resulting in negative mean square radii for these mesons. For example, the one $\langle r^2\rangle_{K^0}=-0.090 ~\text{fm}^2$ slightly deviates from PDG average $-0.077 ~\text{fm}^2$, but agrees with LAI data $-0.090~\text{fm}^2$~\cite{ParticleDataGroup:2022pth}. Consequently, an imaginary value for charge radius is carried out as listed in Table.~\ref{aaa}. This phenomenon is reasonable and is a direct consequence of QCD and the mass hierarchy between the constituent quarks. More explicitly, the mean square charge radius $\left\langle r^2 \right\rangle_P$ is also given by, $\langle r^2 \rangle_P = \int \rho_P(\mathbf{r}) r^2  d^3\mathbf{r}$, where $\rho_P(\mathbf{r})$ is the charge density distribution normalized to the total charge (which is zero for neutral mesons). Then it reflects the spatial distribution of their internal quark charges. For a neutral particle, the positive and negative charges contribute with opposite signs. If the negative charge is more spread out than the positive charge, the integral becomes negative. For mesons such as $K^0 (d\bar{s})$, $D^0 (c\bar{u})$, $B^0 (d\bar{b})$, or $B_s (s\bar{b})$, the heavier $\bar{s}$-antiquark, $c$-quark, or $\bar{b}$-antiquark is localized near the center. Its partner quark or antiquark, which carries an opposite charge, extends farther out. This spatial separation (with the outer charge contributing more to $\left\langle r^2 \right\rangle_P$) leads to a negative mean square charge radius.

\subsection{The heavy flavor mesons}

Furthermore, we proceed to calculate the EMFFs of heavy-flavor mesons. The results have been presented in Fig.~\ref{ff10} and Table~\ref{aaa}. The EMFFs and $\langle r^2\rangle_P$ are closely related to the charge and mass of the heavy (anti)quark (\(m_h\)) as well as the mass difference (\(\Delta m= m_h - m_l\)) between the constituent quarks, where the values of $\Delta m$ are listed in Table~\ref{aaa}. Some findings are as follows:
\begin{itemize}
\item As the $m_h$ increases, the peak values of $Q^2 F_{P}(Q^2)$ also rise significantly, following the order $Q^2 F_{B}>Q^2 F_{D}>Q^2 F_{K}>Q^2 F_{\pi}$. This behavior arises because the explicit breaking of chiral symmetry causes the wave function to localize more strongly, leading to a more compact structure that concentrates the charge distribution. Consequently, photons can couple effectively even at high $Q^2$, resulting in a higher peak of the EMFF and a shift toward higher $Q^2$ values. For the same heavy-flavor hybrid states, such as $B^+$, $B^0$, $B_s$, and $B_c$, $Q^2 F_{P}(Q^2)$ increases remarkably with larger $\Delta m$.

\item For charged mesons, we can find that the charge radius $\sqrt{\langle r^2\rangle_P}$ is negatively correlated with the mass of heavy flavored mesons $m_h$, and positively correlated with the charge of heavy flavored quarks $e_h$ and the mass difference of constituent quarks $\Delta m$ from Table.\ref{aaa}. Notably, the size of meson is also closely related to the parameter $\beta$, our calculations demonstrate that a larger $\beta$ corresponds to a smaller charge radius. Therefore, we attempt to provide a phenomenological relationship  $\sqrt{\langle r^2\rangle_P}\propto e_h \Delta m^l/\beta^{\frac{3}{2}} m_h^s$.

\item As discussion in the subsection~\ref{sec:pika}, the endpoint behaviors of the neutral mesons EMFFs are manifestly differ from the ones of charged mesons, as shown in Fig.\ref{ff10}. The $\sqrt{\langle r^2\rangle}$ of neutral mesons primarily determined by the charge of heavy flavor quark, thus the  $\sqrt{\langle r^2\rangle_{D^0}}$ is the largest one and about two times of the $\sqrt{\langle r^2\rangle_{K^0}}$. In addition, we can see that compared to $D^0$, although the heavy quark charges of $B^0$ and $B_s$ are smaller, they still have sizes comparable to $\sqrt{\langle r^2\rangle_{D^0}}$. This is because the mass asymmetry caused by the large mass difference of constituent quarks $\Delta m$ plays a nontrivial role.
\end{itemize}

\section{CONCLUSIONS}\label{sec:4}

In this work, the EMFFs $F_P(Q^2)$ and charge radii $\sqrt{\langle r^2\rangle_{P}}$ of pseudoscalar mesons are investigated within LFQM. Numerical calculations are performed using parameters constrained by decay constants. Our results are generally in agreement with the currently available experimental data for pions and kaons, and roughly consistent with other theoretical predictions for heavy-flavor $D$- and $B$-mesons.
In addition, the $Q^2$-dependence of the EMFFs and the behavior of charge radii have been analyzed, with the findings as follows: For charged mesons, the peak values of $Q^2 F_P(Q^2)$ are approximately proportional to the mass difference $\Delta m$ between constituent quarks. This proportionality also holds for neutral mesons. The charge radii $\sqrt{\langle r^2 \rangle_P}$ of charged mesons decrease as the meson mass (or heavy-quark mass) increases, reflecting the weakening of relativistic effects in the heavy-quark limit. Additionally, $\Delta m$ plays a nontrivial role in this behavior. For neutral mesons, the heavy-quark charge dominantly determines $\sqrt{\langle r^2 \rangle_P}$. Collectively, these results demonstrate that mass asymmetry significantly influences the EMFFs and charge radii, and these theoretical predictions will be tested in future experiments.

\section*{Acknowledgements}

We thank Qin Chang and Baihui Cheng for many helpful discussions. The work is supported by the National Natural Science Foundation of China (Grant No.11875122, No.12175025 and No.12347101) and Tongling University Talent Program (Grant No.R23100).


\begin{thebibliography}{99}
\bibitem{Dally:1981ur}
E.~B.~Dally, \textit{et al.}
Phys. Rev. D \textbf{24}, 1718 (1981)

\bibitem{Dally:1982zk}
E.~B.~Dally, \textit{et al.}
Phys. Rev. Lett. \textbf{48}, 375 (1982)

\bibitem{Dally:1980dj}
E.~B.~Dally, \textit{et al.}
Phys. Rev. Lett. \textbf{45}, 232 (1980)
\bibitem{Amendolia:1984nz}
S.~R.~Amendolia, \textit{et al.} [NA7],
Phys. Lett. B \textbf{146}, 116 (1984)

\bibitem{NA7:1986vav}
S.~R.~Amendolia, \textit{et al.} [NA7],
Nucl. Phys. B \textbf{277}, 168 (1986)

\bibitem{Amendolia:1986ui}
S.~R.~Amendolia, \textit{et al.}
Phys. Lett. B \textbf{178}, 435 (1986)

\bibitem{Arrington:2021alx}
J.~Arrington, \textit{et al.}
Prog. Part. Nucl. Phys. \textbf{127}, 103985 (2022)

\bibitem{Dudek:2012vr}
J.~Dudek, \textit{et al.}
Eur. Phys. J. A \textbf{48}, 187 (2012)

\bibitem{Bebek:1974ww}
C.~J.~Bebek, \textit{et al.}
Phys. Rev. D \textbf{13}, 25 (1976)

\bibitem{Bebek:1976qm}
C.~J.~Bebek, \textit{et al.}
Phys. Rev. Lett. \textbf{37}, 1326 (1976)

\bibitem{Bebek:1977pe}
C.~J.~Bebek, \textit{et al.}
Phys. Rev. D \textbf{17}, 1693 (1978)

\bibitem{Kunszt:1979ur}
Z.~Kunszt and E.~Pietarinen,
Z. Phys. C \textbf{2}, 355 (1979)

\bibitem{Ackermann:1977rp}
H.~Ackermann, \textit{et al.}
Nucl. Phys. B \textbf{137}, 294 (1978)

\bibitem{JeffersonLabFpi:2000nlc}
J.~Volmer \textit{et al.} [Jefferson Lab F(pi)],
Phys. Rev. Lett. \textbf{86}, 1713 (2001)


\bibitem{JeffersonLabFpi-2:2006ysh}
T.~Horn \textit{et al.} [Jefferson Lab F(pi)-2],
Phys. Rev. Lett. \textbf{97}, 192001 (2006)


\bibitem{JeffersonLabFpi:2007vir}
V.~Tadevosyan \textit{et al.} [Jefferson Lab F(pi)],
Phys. Rev. C \textbf{75}, 055205 (2007)

\bibitem{JeffersonLab:2008gyl}
H.~P.~Blok \textit{et al.} [Jefferson Lab],
Phys. Rev. C \textbf{78}, 045202 (2008)

\bibitem{JeffersonLab:2008jve}
G.~M.~Huber \textit{et al.} [Jefferson Lab],
Phys. Rev. C \textbf{78}, 045203 (2008)

\bibitem{Carmignotto:2018uqj}
M.~Carmignotto, \textit{et al.}
Phys. Rev. C \textbf{97}, 025204 (2018)

\bibitem{AbdulKhalek:2021gbh}
R.~Abdul Khalek, \textit{et al.}
Nucl. Phys. A \textbf{1026}, 122447 (2022)

\bibitem{Li:2017eic}
N.~Li and Y.~J.~Wu,
Eur. Phys. J. A \textbf{53}, 56 (2017)


\bibitem{Ding:2024lfj}
H.~T.~Ding, X.~Gao, A.~D.~Hanlon, S.~Mukherjee, P.~Petreczky, Q.~Shi, S.~Syritsyn, R.~Zhang and Y.~Zhao,
Phys. Rev. Lett. \textbf{133}, 181902 (2024)

\bibitem{Nesterenko:1982gc}
V.~A.~Nesterenko and A.~V.~Radyushkin,
Phys. Lett. B \textbf{115}, 410 (1982)


\bibitem{Roberts:1994dr}
C.~D.~Roberts and A.~G.~Williams,
Prog. Part. Nucl. Phys. \textbf{33}, 477 (1994)

\bibitem{Maris:1999nt}
P.~Maris and P.~C.~Tandy,
Phys. Rev. C \textbf{60}, 055214 (1999)

\bibitem{Qin:2011dd}
S.~X.~Qin, L.~Chang, Y.~X.~Liu, C.~D.~Roberts and D.~J.~Wilson,
Phys. Rev. C \textbf{84}, 042202 (2011)

\bibitem{Xu:2019ilh}
Y.~Z.~Xu, D.~Binosi, Z.~F.~Cui, B.~L.~Li, C.~D.~Roberts, S.~S.~Xu and H.~S.~Zong,
Phys. Rev. D \textbf{100}, 114038 (2019)

\bibitem{Xu:2020loz}
Y.~Z.~Xu, C.~Shi, X.~T.~He and H.~S.~Zong,
Phys. Rev. D \textbf{102}, 114011 (2020)

\bibitem{Maris:1997tm}
P.~Maris and C.~D.~Roberts,
Phys. Rev. C \textbf{56}, 3369 (1997)

\bibitem{Qin:2019oar}
P.~Qin, S.~X.~Qin and Y.~X.~Liu,
Phys. Rev. D \textbf{101}, 114014 (2020)

\bibitem{Xu:2024fun}
Y.~Z.~Xu,
JHEP \textbf{07}, 118 (2024)

\bibitem{Moita:2021xcd}
R.~M.~Moita, J.~P.~B.~C.~de Melo, K.~Tsushima and T.~Frederico,
Phys. Rev. D \textbf{104}, 096020 (2021)

\bibitem{Luan:2015goa}
Y.~L.~Luan, X.~L.~Chen and W.~Z.~Deng,
Chin. Phys. C \textbf{39}, 113103 (2015)

\bibitem{Hernandez-Pinto:2023yin}
R.~J.~Hern{\'a}ndez-Pinto, L.~X.~Guti{\'e}rrez-Guerrero, A.~Bashir, M.~A.~Bedolla and I.~M.~Higuera-Angulo,
Phys. Rev. D \textbf{107}, 054002 (2023)

\bibitem{Almeida-Zamora:2023bqb}
B.~Almeida-Zamora, J.~J.~Cobos-Mart{\'\i}nez, A.~Bashir, K.~Raya, J.~Rodr{\'\i}guez-Quintero and J.~Segovia,
Phys. Rev. D \textbf{109}, 014016 (2024)

\bibitem{Huang:2004fn}
T.~Huang, X.~G.~Wu and X.~H.~Wu,
  Phys. Rev. D \textbf{70}, 053007 (2004)

\bibitem{Huang:2004su}
T.~Huang and X.~G.~Wu,
Phys. Rev. D \textbf{70}, 093013 (2004)

\bibitem{Wu:2008yr}
X.~G.~Wu and T.~Huang,
JHEP \textbf{04}, 043 (2008)


\bibitem{Wang:2019mhm}
B.~Wang, B.~Yang, L.~Meng and S.~L.~Zhu,
Phys. Rev. D \textbf{100}, 016019 (2019)

\bibitem{Das:2016rio}
T.~Das, D.~K.~Choudhury and N.~S.~Bordoloi,
[arXiv:1608.06896 [hep-ph]]

\bibitem{Aliev:2019lsd}
T.~M.~Aliev, S.~Bilmis and M.~Savci,
Phys. Rev. D \textbf{101}, 054009 (2020)

\bibitem{Simonis:2016pnh}
V.~Simonis,
Eur. Phys. J. A \textbf{52}, 90 (2016)

\bibitem{Bose:1980vy}
S.~K.~Bose and L.~P.~Singh,
Phys. Rev. D \textbf{22}, 773 (1980)

\bibitem{Lahde:2002wj}
T.~A.~Lahde,
Nucl. Phys. A \textbf{714}, 183 (2003)

\bibitem{Brodsky:1992px}
  S.~J.~Brodsky and J.~R.~Hiller,
  Phys.\ Rev.\ D {\bf 46}, 2141 (1992)

\bibitem{Jaus:1999zv}
  W.~Jaus,
  Phys.\ Rev.\ D {\bf 60}, 054026 (1999)

\bibitem{Grach:1983hd}
  I.~L.~Grach and L.~A.~Kondratyuk,
  Sov.\ J.\ Nucl.\ Phys.\  {\bf 39}, 198 (1984)

\bibitem{Chung:1988my}
  P.~L.~Chung, W.~N.~Polyzou, F.~Coester and B.~D.~Keister,
  Phys.\ Rev.\ C {\bf 37}, 2000 (1988)

\bibitem{Hwang:2001th}
C.~W.~Hwang,
Eur. Phys. J. C \textbf{23}, 585 (2002)

\bibitem{Chang:2019obq}
  Q.~Chang, L.~T.~Wang and X.~N.~Li,
  JHEP {\bf 1912}, 102 (2019).

\bibitem{Chang:2018zjq}
  Q.~Chang, X.~N.~Li, X.~Q.~Li, F.~Su and Y.~D.~Yang,
  Phys.\ Rev.\ D {\bf 98}, 114018 (2018)

\bibitem{Chang:2019mmh}
  Q.~Chang, X.~N.~Li and L.~T.~Wang,
  Eur.\ Phys.\ J.\ C {\bf 79}, 422 (2019)

\bibitem{Jaus:1989au}
  W.~Jaus,
  Phys.\ Rev.\ D {\bf 41}, 3394 (1990).

\bibitem{Jaus:1996np}
  W.~Jaus,
  Phys.\ Rev.\ D {\bf 53}, 1349 (1996)

\bibitem{Cheng:2003sm}
  H.~Y.~Cheng, C.~K.~Chua and C.~W.~Hwang,
  Phys.\ Rev.\ D {\bf 69},  074025 (2004)

 \bibitem{Choi:2013mda}
  H.~M.~Choi and C.~R.~Ji,
  Phys.\ Rev.\ D {\bf 89}, 033011 (2014)

\bibitem{S:2024adt}
T.~M.~S., A.~Hazra, N.~Sharma and R.~Dhir,
Eur. Phys. J. C \textbf{85}, 204 (2025)

\bibitem{Choi:2017uos}
  H.~M.~Choi and C.~R.~Ji,
  Phys.\ Rev.\ D {\bf 95}, 056002 (2017)
\bibitem{Xu:2025aow}
S.~Xu, X.~N.~Li and X.~G.~Wu,
[arXiv:2505.02419 [hep-ph]].
\bibitem{Choi:1997iq}
H.~M.~Choi and C.~R.~Ji,
Phys. Rev. D \textbf{59}, 074015 (1999)



\bibitem{Choi:1999nu}
H.~M.~Choi and C.~R.~Ji,
Phys. Lett. B \textbf{460}, 461 (1999)


\bibitem{Choi:2009ai}
H.~M.~Choi and C.~R.~Ji,
Phys. Rev. D \textbf{80}, 054016 (2009)

\bibitem{Hwang:2010hw}
C.~W.~Hwang,
Phys. Rev. D \textbf{81}, 114024 (2010)

\bibitem{Verma:2011yw}
R.~C.~Verma,
J. Phys. G \textbf{39}, 025005 (2012)

\bibitem{Choi:2015ywa}
H.~M.~Choi, C.~R.~Ji, Z.~Li and H.~Y.~Ryu,
Phys. Rev. C \textbf{92}, 055203 (2015)


\bibitem{ParticleDataGroup:2022pth}
R.~L.~Workman \textit{et al.} [Particle Data Group],
PTEP \textbf{2022}, 083C01 (2022)

\bibitem{Lepage:1979zb}
G.~P.~Lepage and S.~J.~Brodsky,
Phys. Lett. B \textbf{87}, 359 (1979)

\end{thebibliography}
\end{document}